\documentstyle[seceq,epsf,mbf]{ptptex}

\title{
General Relativistic Modification of a 
Pulsar Electromagnetic Field
}

\author{
Kohkichi {\sc Konno}\footnote{
E-mail: konno@theo.phys.sci.hiroshima-u.ac.jp} 
and Yasufumi {\sc Kojima}\footnote{
E-mail: kojima@theo.phys.sci.hiroshima-u.ac.jp}
}

\inst{
Department of Physics, Hiroshima University,
Higashi-Hiroshima 739-8526, Japan
}

\recdate{
September 4, 2000
}

\abst{
 We consider an exterior electromagnetic field surrounding a rotating 
 star endowed with a dipole magnetic field in the context 
 of general relativity. 
 The analytic solution for a stationary configuration
 is obtained, and the general relativistic modifications
 and the implications for pulsar radiation are investigated
 in detail. We find that the general relativistic corrections
 of both the electric field strength and the curvature radii
 of magnetic field lines tend to enhance the
 curvature radiation photon energy.
}

\begin{document}

\maketitle

\section{Introduction}

In recent years, new aspects of rotating neutron stars
have been revealed in about 1000 pulsars.
Eleven X-ray pulsars\cite{bt99} and 
eight $\gamma$-ray pulsars\cite{th99} have been detected 
in the past several years.
Among these new objects, some exhibit quite different behavior
in their pulse periods.\cite{obs}
The measurement of the period and its time derivative
yields evidence of ultra-magnetized stars,
possibly representing magnetars.\cite{dt}
Motivated by the recent observational situation, 
theoretical models have been studied.
As for high-energy pulsars,
two general classes of models have been proposed.
One is the polar cap model\cite{pol}
and the other is the outer gap model.\cite{out}
The main difference between these two models is in the 
assumed region of the acceleration of charged particles
responsible for the radiation.
Both models partially explain 
some observational features of the  $\gamma$-rays.
They will be discriminated after
including more detailed radiation processes.
Future observation may determine their validity.

An important element to be included in theoretical models
is general relativistic effects,
which are in particular crucial for polar cap models,
since acceleration occurs under strong gravity near
the surface of neutron stars.
Gonthier and Harding\cite{gh94} considered
the effects on the magnetic field configuration only.
Their concern is the curvature radiation and
the attenuation of pair production in a strong magnetic field.
These processes result in a pair cascade and 
explain some aspects of pulsar radiation, including 
high-energy pulses in the $\gamma$-ray range.
In addition to the magnetic fields,
rotationally induced electric fields play an important 
role in the polar cap region (see, e.g., Ref.~\citen{meszaros}).
Charged particles are ripped off 
the surface and accelerated along the magnetic field lines
by the electric fields.
The magnetosphere is thereby eventually filled with charges.
The accelerated particles may be seeds of subsequent 
curvature radiation.
Muslimov and Tsygan\cite{mt} discussed 
general relativistic effects not only in the case of magnetic 
fields but also electric fields.
They derived general expressions including multipoles
of arbitrary order using
hypergeometric functions, assuming a vacuum outside the star.
Their work, however, is limited to only analytic forms, 
and therefore not easy, e.g., to compare with 
the standard results in flat space-time.
Order estimates of general relativistic effects
are also lacking.
In this paper, we derive analytic 
solutions again for both the electric and magnetic
fields around a rotating star endowed with an aligned dipole
magnetic field. The resultant expressions are rather cumbersome, 
and for this reason approximate expressions are also given.
Such forms provide an estimate of 
the corrections to the results in flat space-time,
as well as a concise, practical application.
We also give detailed discussion concerning the 
difference between our results and those
in Minkowski space-time.
This discussion may become important in the future,
with progress in observational technology.

As shown in Ref.~\citen{kok2}, 
the deviation from spherical space-time
is less than $ 10^{-3}$ 
if the rotation period is longer than 10 msec
and the magnetic field at the surface is
less than  $ 10^{16}$ gauss.
Therefore, the electric and magnetic fields
are determined by solving the Maxwell equations
in a fixed background space-time.
The appropriate space-time metric is that for an external
field surrounding a slowly rotating star.
We can neglect the second-order rotational effects,
except in the case of rapidly rotating stars.  
We also restrict ourselves to a stationary configuration,
that is, the case in which the magnetic dipole moment 
$\mib{\mu}$ is aligned with the 
angular velocity $\mib{\Omega}$.
This leads to the following form $A_{\mu} = (A_{t},0,0,A_{\phi})$
for the four-potential 
(see Ref.~\citen{bgsm} and references therein),
where $A_{t}$ is related to the rotationally induced 
electric field, and therefore   
$A_{t} \sim O \left(  \Omega \right)\times  A_{\phi}  $.
Detailed calculations to solve the Maxwell equations 
are given in \S 2.
Approximate expressions of these solutions
are discussed in \S 3. Implications of the 
general relativistic effects with regard to 
the acceleration of charged 
particles and radiation in vacuum gaps 
are investigated in \S 4. Finally, we give discussion in \S 5.
Throughout the paper, we use units in which $c=G=1$.

\section{The general relativistic solution for an 
         exterior stellar electromagnetic field}

We now derive expressions for an electromagnetic field 
surrounding a rotating, magnetized star using a
general relativistic treatment.
We solve the Maxwell equations in a fixed metric,
assuming that the field is in a vacuum. 
The background metric outside the star
with total mass $M$ and angular momentum $J$ 
is specified up to first order in the slow rotation approximation as
\begin{equation}
 ds^2 = - e^{-\lambda (r) } dt^2 
        - 2 \omega (r) r^2 \sin^2 \theta dt d\phi
        + e^{\lambda (r)} dr^2 
        + r^2 d\theta^2 + r^2 \sin^2 \theta d\phi^2 ,
\end{equation}
where
\begin{eqnarray}
e^{\lambda} &=& 
 \left( 1 - \frac{2M}{r} \right)^{-1},
\\
\omega & =& \frac{2J }{ r^3 }.
\end{eqnarray}

In the non-rotating limit, a
poloidal  magnetic field can be described by the
$A_{\phi}$ component only. 
In the slowly  rotating case,   
the four-potential is given by 
$A_{\mu} = (A_{t},0,0,A_{\phi}).$ 
The $ A_{t} $ component is rotationally induced as
$ A_{t} \sim  O(\Omega )  \times  A_{\phi}$.
The Maxwell equations for $A_{t}$ and $A_{\phi}$
are given as
\begin{subequations}
\begin{equation}
\label{eq.ap}
 e^{-\lambda} \frac{\partial^2 A_{\phi}}{\partial r^2}
 - \lambda' e^{-\lambda}
   \frac{\partial A_{\phi}}{\partial r}
 + \frac{1}{r^2} \frac{\partial^2 A_{\phi}}{\partial \theta^2}
 - \frac{1}{r^2} \cot \theta \frac{\partial A_{\phi}}
   {\partial \theta} = 0 ,
\end{equation}
\begin{eqnarray}
\label{eq.at}
 e^{-\lambda} \frac{\partial^2 A_{t}}{\partial r^2}
   + \frac{2 e^{-\lambda}}{r} \frac{\partial A_{t}}{\partial r}
  + \frac{1}{r^2} \frac{\partial^2 A_{t}}{\partial \theta^2}
  + \frac{1}{r^2} \cot \theta 
  \frac{\partial A_{t}}{\partial \theta} & & \nonumber \\
 + \left[ \left( \lambda' + \frac{2}{r} \right) \omega 
  + \omega' \right] e^{-\lambda} \frac{\partial A_{\phi}}{\partial r}
  + \frac{2}{r^2} \omega \cot \theta 
  \frac{\partial A_{\phi}}{\partial \theta}
 & = & 0 ,
\end{eqnarray}
\end{subequations}
where the prime here denotes 
differentiation with respect to $r$.
Note that 
the last two terms on the left-hand side
of Eq.~(\ref{eq.at}) represent
the coupling between the frame-dragging 
and the stellar magnetic field, and that these terms 
originate from a purely general relativistic effect.

From this point, we restrict our discussion to 
the case of a dipole magnetic field, 
so that Eq.~(\ref{eq.ap}) can be solved in the form
\begin{equation}
\label{ap_exp}
 A_{\phi}(r,\theta ) = - a_{\phi}(r) \sin^2 \theta .
\end{equation}
In a similar way, 
the potential $A_{t}$ can be written as
\begin{equation}
 A_{t}(r,\theta ) = a_{t0}(r) + a_{t2}(r) 
   P_{2} \left( \cos \theta \right) ,
\end{equation}
where $P_{2}$ is the Legendre polynomial of degree 2.

The solution for $a_{\phi}$ can easily be derived in the form\cite{go}
\begin{equation}
 a_{\phi} = \frac{3\mu}{8M^3} r^2 \left[ \ln \left( 
  1 - \frac{2M}{r} \right) + \frac{2M}{r} 
  + \frac{2M^2}{r^2}\right] ,
\end{equation}
where $\mu$ is the magnetic dipole moment with respect to
an observer at infinity. 
The resulting dipole magnetic field in the local frame is 
given by
\begin{subequations}
\label{s-b}
\begin{eqnarray}
\label{b1}
 B_{(r)} & = & - \frac{3\mu}{4M^3} 
  \left[ \ln \left( 1 - \frac{2M}{r} \right) + \frac{2M}{r} 
  + \frac{2M^2}{r^2}\right] \cos \theta , \\
\label{b2}
 B_{( \theta )} & = & \frac{3\mu}{4M^3} \left[
  \sqrt{1-\frac{2M}{r}} \ln \left( 1 - \frac{2M}{r} \right) 
  + \frac{2M(r-M)}{r\sqrt{r(r-2M)}} \right] \sin \theta .
\end{eqnarray}
\end{subequations}

Next, 
we discuss the electric field induced by
the rigid rotation of the star. 
The solution for $a_{t0}$ and $a_{t2}$ can be obtained analytically as
\begin{subequations}
\begin{eqnarray}
 a_{t0} & = & \frac{c_{0}}{r} + \frac{J\mu}{2M^3r^2}
   (3r-M) + \frac{J\mu}{4M^4r} (3r-4M) \ln 
   \left( 1 - \frac{2M}{r} \right) , \\
 a_{t2} & = & \frac{c_{1}}{M^2} (r-M)(r-2M) \nonumber \\
 & & + c_{2} \left[ \frac{2}{Mr} \left( 3r^2-6Mr+M^2 \right)
     + \frac{3}{M^2} \left( r^2-3Mr+2M^2 \right)
     \ln \left( 1 - \frac{2M}{r} \right)\right] \nonumber \\
 & & \qquad - \frac{J\mu}{2M^6r^2} \left( 9r^4-3Mr^3
     -30M^2r^2+8M^3r+2M^4 \right) \nonumber \\
 & & \qquad - \frac{J\mu}{2M^6r} \left( 12r^3-36Mr^2
     +24M^2r+M^3 \right) \ln
     \left( 1 - \frac{2M}{r} \right) ,
\end{eqnarray}
\end{subequations}
where $c_{0}$, $c_{1}$ and $c_{2}$ are constants of integration.
Since $c_{0}$ is understood as the net charge of the star,
we set $c_{0}=0$. Furthermore, we derive
\begin{equation}
 c_{1} = \frac{9J\mu}{2M^4} ,
\end{equation} 
from the regularity condition at infinity.
The constant $c_{2}$ is fixed by the junction condition 
at the surface
of the star. If we impose the assumption of a perfectly 
conducting interior, 
the magnetic field is frozen into the the fluid motion,
i.e.~$u^{\mu} F_{\mu \nu} = 0$, where
$u^{\mu}= ( u^t ,0,0, \Omega u^t)$ 
is the four-velocity of the fluid.
From this condition at the surface, we have
\begin{eqnarray}
 c_{2} & = & \left\{ \frac{\mu J}{M^5 R^2} 
  \left( 12 R^3 - 24 MR^2 + 4 M^2 R + M^3 \right) \right. \nonumber \\
 & & \quad + \frac{\mu J}{2 M^6 R} \left( 12 R^3 - 36 MR^2 
     + 24 M^2 R + M^3 \right)
     \log \left( 1 - \frac{2M}{R} \right) \nonumber \\
 & & \quad \left. - \frac{\mu \Omega}
  {4M^3} \left[ 2MR + 2M^2 + R^2 \log \left( 1 - \frac{2M}{R} \right)
  \right] \right\} \nonumber \\
 & & \Big/ \left[ \frac{2}{MR} 
  \left( 3R^2 - 6MR + M^2 \right) \right. \nonumber \\
 & & \qquad \left. + \frac{3}{M^2} 
  \left( R^2 - 3MR + 2M^2 \right) \log \left( 1 - \frac{2M}{R} \right)
  \right] , 
\end{eqnarray}
where $R$ denotes the radius of the star.
Consequently, using the above $c_2$, the induced electric field 
in the local frame can be written as
\begin{subequations}
\label{s-e}
\begin{eqnarray}
\label{e1}
 E_{(r)} & = & \frac{1}{2M^6r^3} \left\{ c_{2} \left[
  4M^5r \left( 6r^2-3Mr-M^2 \right) \right. \right. \nonumber \\
 & & \qquad \qquad \left. + 6M^4r^3
  \left( 2r-3M \right) \ln \left( 1 - \frac{2M}{r} \right)
  \right] \nonumber \\
 & & \qquad \qquad - 2MJ\mu \left( 24r^3-12Mr^2-4M^2r-3M^3 \right) 
  \nonumber \\
 & & \qquad \qquad \left. - 3rJ\mu \left( 8r^3-12Mr^2-M^3 \right) 
  \ln \left( 1 - \frac{2M}{r} \right) \right\}
  P_{2} (\cos \theta ) , \nonumber \\ \\
\label{e2}
 E_{(\theta )} & = & - \frac{3}{M^6r^3 \sqrt{r(r-2M)}} \nonumber \\
 & & \times \left\{ c_{2} \left[ 2M^5r^2 
     \left( 3r^2-6Mr+M^2 \right) \right. \right. \nonumber \\
 & & \qquad \left. + 3M^4r^3 
     \left( r^2-3Mr+2M^2 \right) \ln \left( 1 - 
     \frac{2M}{r} \right) \right] \nonumber \\
 & & \qquad - MJ\mu \left( 12r^4-24Mr^3+4M^2r^2-M^4 \right)
     \nonumber \\
 & & \qquad \left. 
     -6r^3J\mu \left( r^2-3Mr+2M^2 \right) \ln
     \left( 1-\frac{2M}{r} \right) \right\} \sin \theta
     \cos \theta .
\end{eqnarray}
\end{subequations}
The discussion of the quantitative nature of 
the electromagnetic field
strength is given in the next section.

\section{Comparison with results in flat space-time}

In the previous section, we obtained expressions 
for the electromagnetic field 
in the general relativistic framework. 
However, these expressions are somewhat cumbersome. 
They are reduced to standard expressions given
in textbooks\cite{meszaros} 
when the gravitational terms are neglected, 
i.e.~when we take the limits $ M, J \to 0 .$
It is important to compare our expressions with
the standard expressions and to examine the differences.
From this point of view, we now derive the simpler approximate
expressions by expanding in powers of $1/r$. 
The lowest-order forms give the 
standard results, and the next terms give their corrections. 
As an approximation, we use the radius $r $ 
in the Schwarzschild coordinates
as the radius in the flat space-time. 
The magnetic and electric fields can be expanded
in the forms
\begin{subequations}
\begin{eqnarray}
\label{pn-br}
 B_{(r)} & \simeq & \frac{2\mu}{r^3}
\left[ 1 +\frac{3 M}{2r}  \right] \cos \theta , \\
\label{pn-bt}
 B_{(\theta)} & \simeq & \frac{\mu}{r^3}
\left[ 
 1 +\frac{2 M}{r} \right] \sin \theta  ,
\end{eqnarray}
\end{subequations}
\begin{subequations}
\label{exp-e}
\begin{eqnarray}
\label{pn-er}
 E_{(r)} & \simeq & - \frac{2\mu R^2 \Omega}{r^4}
\left[  1-\left( \frac{1}{2}-\frac{8R}{3r}\right)\frac{M}{R}
 + \left(1-\frac{2R}{r}\right)\frac{I}{R^3} 
\right] P_{2}\left( \cos \theta \right), \\
\label{pn-et}
 E_{( \theta )} & \simeq & - \frac{2\mu R^2 \Omega}{r^4}
\left[ 1-\left(\frac{1}{6}-\frac{R}{r}\right)\frac{M}{R}
 + \left(1-\frac{3R}{r}\right)\frac{I}{R^3} 
\right] \sin \theta \cos \theta , 
\end{eqnarray}
\end{subequations}
where the terms following the first ones in each of the
square brackets 
are the first-order corrections due to the curved space-time.
In Eq.(\ref{exp-e}), 
the moment of the inertia $I = J/\Omega $ is used.  
These corrections can be estimated easily for stars 
with uniform density, in which 
$I \sim  2 M R^2/5 $ and $ M/R \le 4/9. $
The correction terms become larger with the
relativistic factor $ M/R, $
but they are less than $1.$
Thus we see that the expressions obtained
in the flat space-time are accurate to within 
a factor of 2.

\begin{figure}
  \epsfxsize = 8cm
  \centerline{\epsfbox{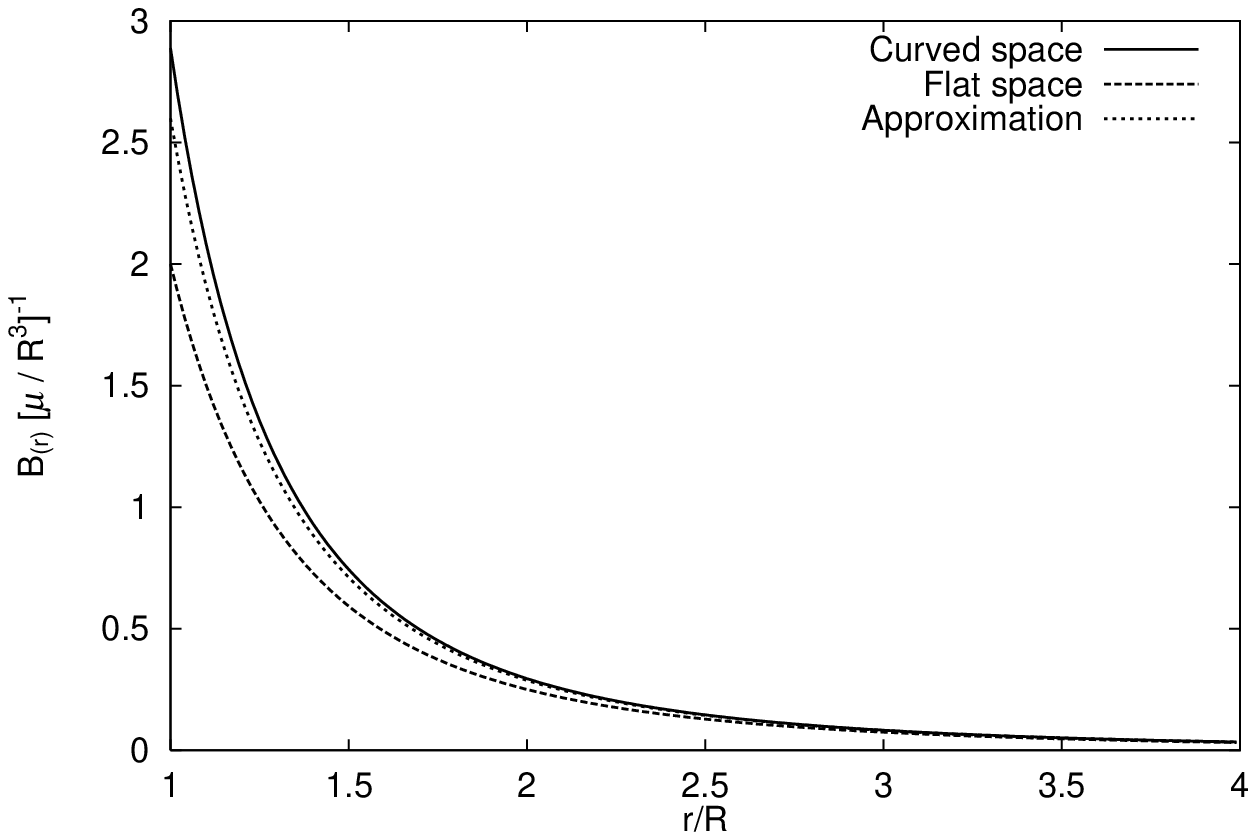}}
  \epsfxsize = 8cm
  \centerline{\epsfbox{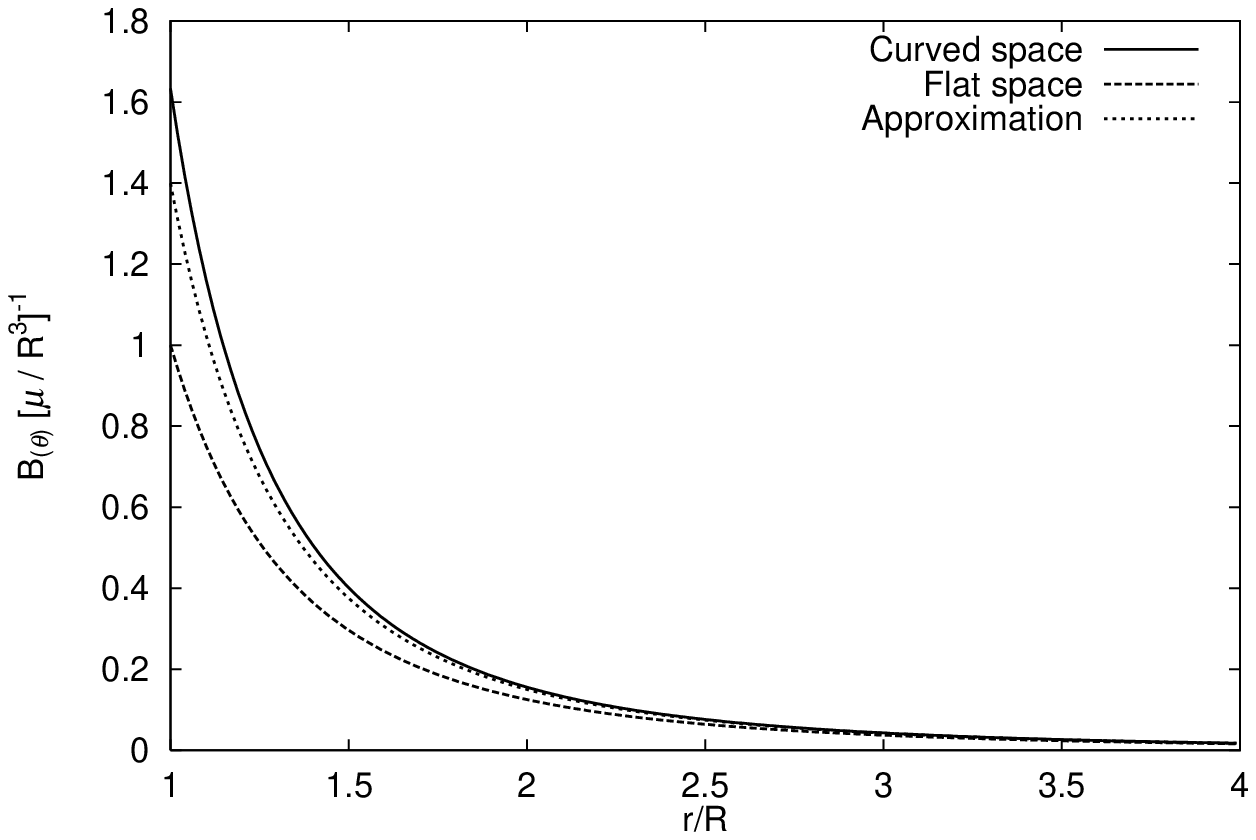}}
  \caption{Radial parts of the magnetic field components
       $B_{(r)}$ and $B_{(\theta)}$ are plotted as functions
       of the radius. The field strength is normalized by the 
       typical value $\mu / R^3$. 
       The solid, dashed and dotted curves denote 
       the curved space-time, flat space-time 
       and approximate expressions, respectively.}
  \label{fig-pn1}
\end{figure}

\begin{figure}
  \epsfxsize = 8cm
  \centerline{\epsfbox{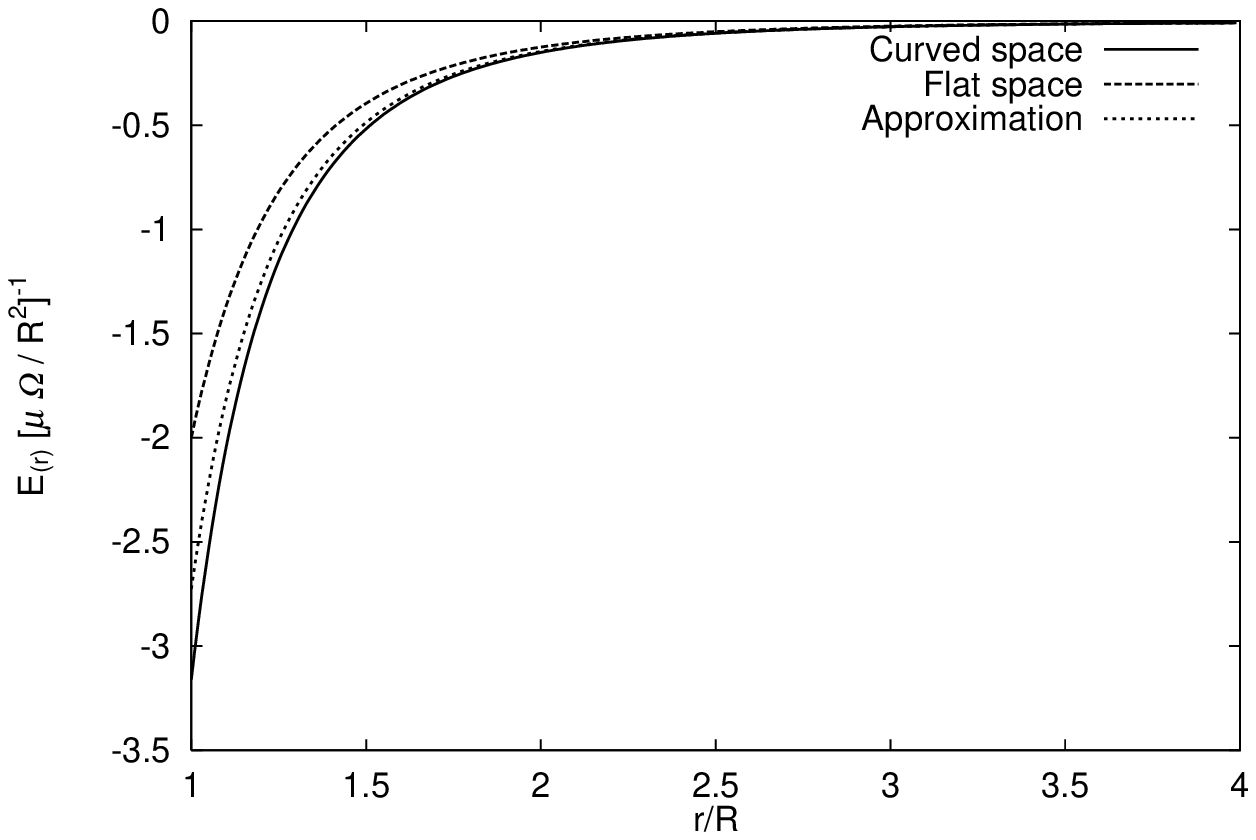}}
  \epsfxsize = 8cm
  \centerline{\epsfbox{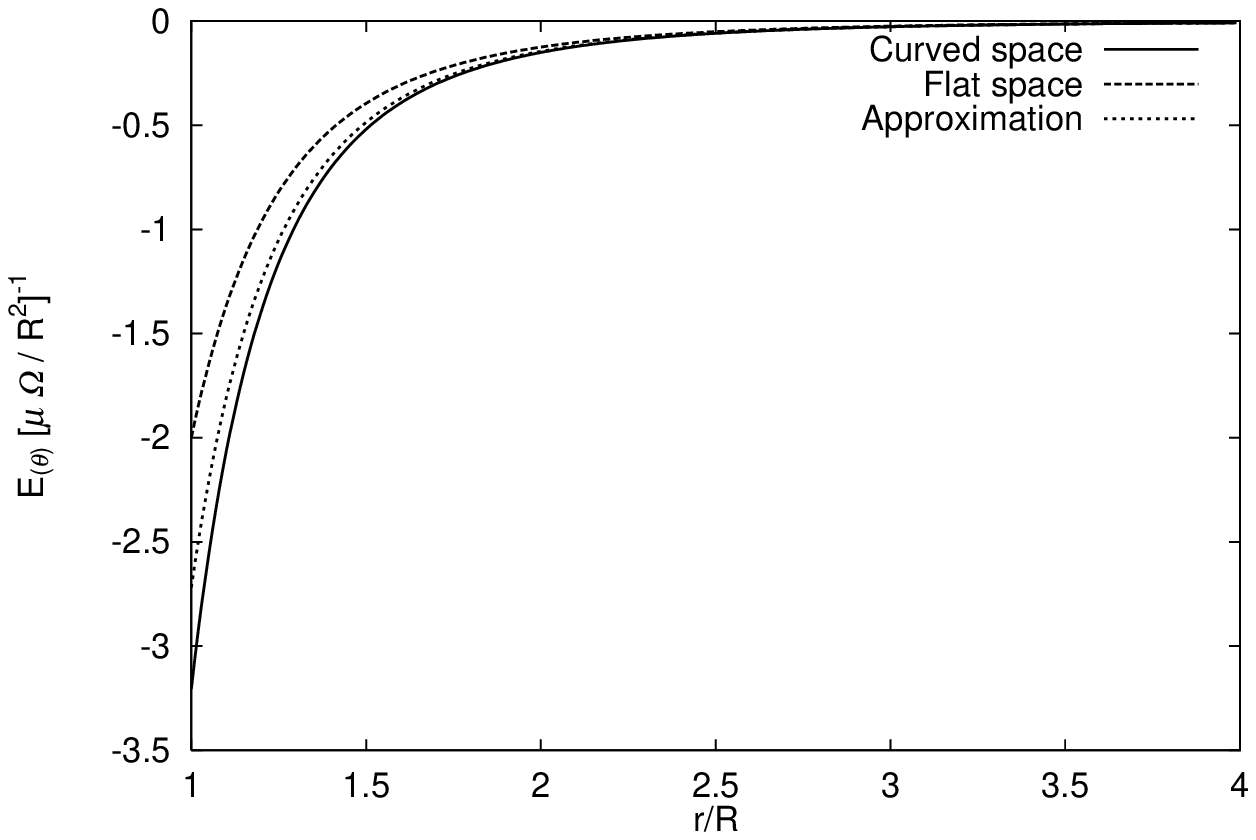}}
  \caption{Radial parts of the magnetic field components
       $E_{(r)}$ and $E_{(\theta)}$ are plotted as functions 
       of the radius. The field strength is normalized by the 
       typical value $\mu \Omega / R^2$. 
       The solid, dashed and dotted curves denote 
       the curved space-time, flat space-time 
       and approximate expressions, respectively.}
  \label{fig-pn2}
\end{figure}

In Figs.~\ref{fig-pn1} and \ref{fig-pn2},
we explicitly display the results
in the flat and curved space-times as functions of the radius.
These figures display the normalized values 
of the radial parts of the magnetic and electric fields, respectively.
We have adopted a polytropic stellar model with $M/R=0.2$, 
which is a plausible value for neutron stars.
The solid curves here denote the exact values in the
curved space-time, 
while the dashed curves correspond to the standard results
in the flat space-time.
From these figures, we find that the standard
expressions in the flat space-time
give values deviating from the curved space-time values  
by 50\% at most. 
The maximum error is roughly estimated as $ 2M/r$. 
Therefore, the standard expressions are 
useful for arguments within 
this order of the magnitude.

\section{Implications for the acceleration of charged particles 
         and the radiation in vacuum gaps}

In this section, 
the results for the electromagnetic field
in curved space-time are applied to analysis of the
pulsar emission mechanism, 
that is, quantities relevant to the 
acceleration of charged particles and radiation 
in vacuum gaps above the polar caps.
The gravitational force is much less than
the electrostatic force, but gravity
affects space-time, whose effects on the
electromagnetic field are considered here.
We explicitly derive the electric field along the magnetic field lines,
curvature radii of the field lines, and size of polar cap regions.
They are important to evaluate the available potential energy, 
curvature radiation, and so on. 
They significantly depend on the
global shape of  magnetic field lines, 
so that deviation from the standard results in 
flat space-time is not estimated using some local positions,
although the overall error is not expected to be large.

\begin{figure}
  \epsfxsize = 8cm
  \centerline{\epsfbox{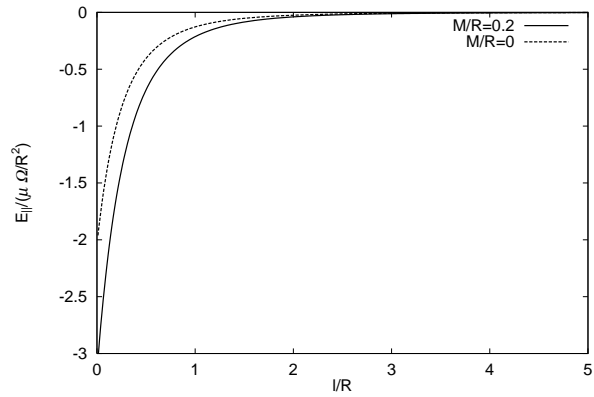}}
  \caption{The electric field component along a magnetic field 
     line that flows from the stellar surface with 
     $\theta = 1^{\circ}$. The field strength, which 
     is normalized by the typical value $\mu \Omega / R^2$,
     is calculated for the Minkowskian case $M/R=0$ (dashed) and
     the relativistic case $M/R=0.2$ (solid). The proper distance
     $l$ from the stellar surface is normalized by $R$.}
  \label{fig4}
\end{figure}

First, we investigate the electric field component
along the magnetic field lines. This plays a direct,
important role in the acceleration of charged particles.
The component is derived from 
Eqs.~(\ref{s-b}) and (\ref{s-e}) as
\begin{equation}
 E_{||} = \frac{E_{(r)} B{_{(r)}} + E_{(\theta)} B_{(\theta)}}
       {\sqrt{B_{(r)}^2 + B_{(\theta)}^2}} .
\end{equation}
Figure \ref{fig4} displays $E_{||}$ normalized by the typical
value $\mu \Omega / R^2$ as a function of the proper distance $l$
from the stellar surface along a field line. 
The dashed curve denotes the Minkowskian case, and 
the solid curve denotes the general relativistic 
case of $M/R=0.2$. This figure shows that 
the electric field component is strengthened by the 
general relativistic effect with respect to the same 
value of $\mu \Omega / R^2$.
The result in the curved case
is about 1.5 times as large as that in the flat case
near the surface.
A similar kind of enhancement 
can be seen in the stellar interior
due to the general relativistic effect.\cite{kok2}
These enhancements may be regarded as
having a common origin.

\begin{figure}
  \epsfxsize = 8cm
  \centerline{\epsfbox{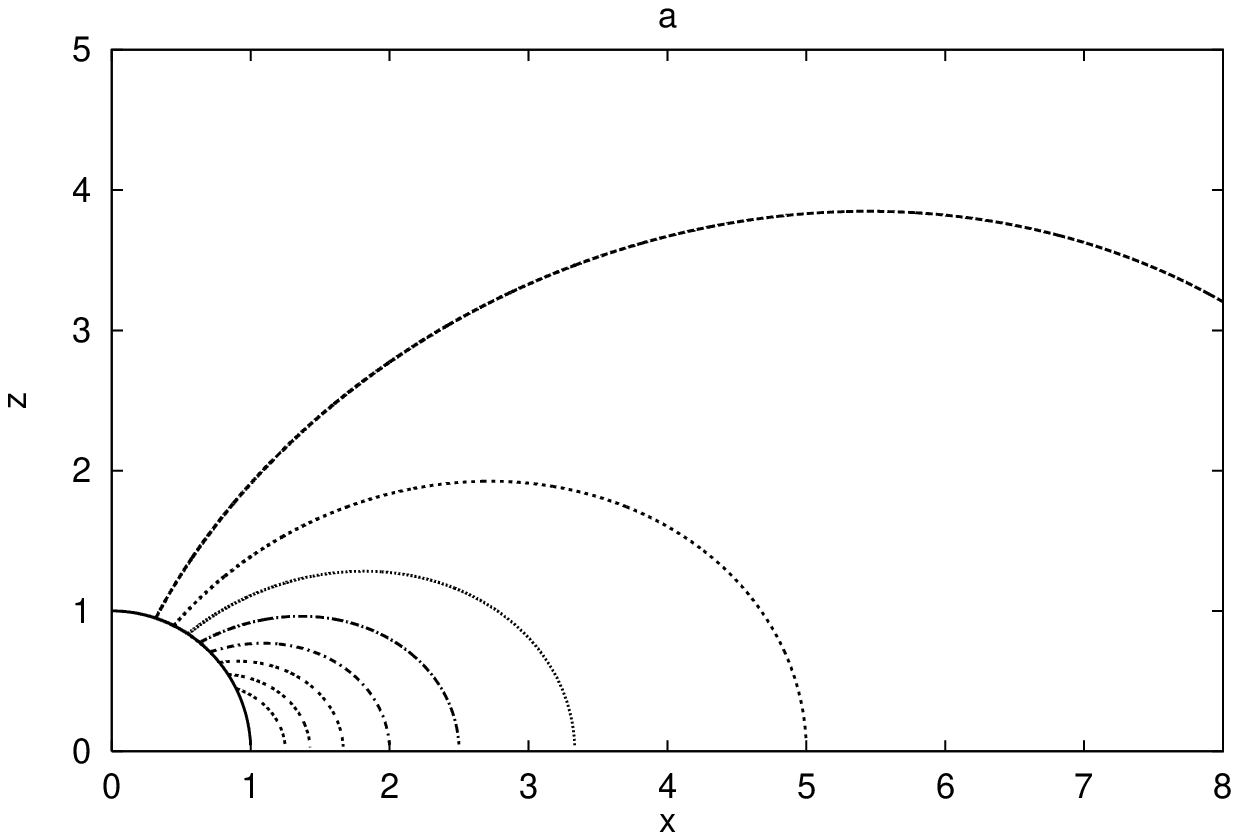}}
  \epsfxsize = 8cm
  \centerline{\epsfbox{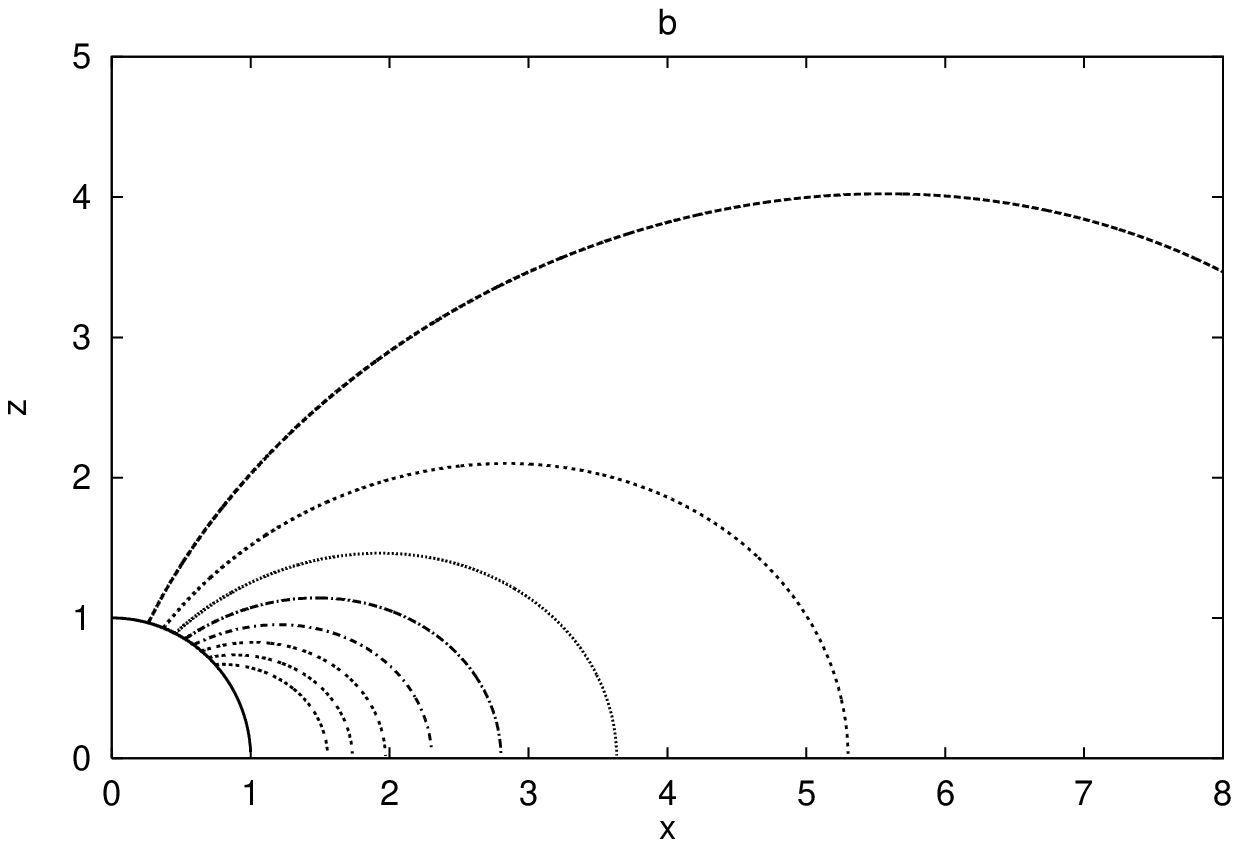}}
  \caption{Magnetic field lines for the Minkowskian case ($M/R=0$) 
      (a) and the general relativistic case 
      with $M/R=0.2$ (b),
      plotted in the $z$-$x$ plane,
      where $(z,x) = (r \sin \theta, r \cos \theta)$.
      Both cases have the same magnetic moment.
      The surface of the star is denoted by the circle of radius 1.}
  \label{fig1}
\end{figure}

The configurations 
of the magnetic field lines are also modified by the general
relativistic effect. 
In general, a magnetic field line is described by
an ordinary differential equation:\cite{dkl}
\begin{equation}
 \frac{dr}{d\theta} = \frac{B_{r}}{B_{\theta}} .
\end{equation}
The solution of this equation is
\begin{equation}
 \label{fl-traj}
 A_{\phi} = \mbox{const} \left( \equiv \tilde{c} \right) .
\end{equation}
Each field line is labeled by a constant $  \tilde{c} . $
Figure \ref{fig1} displays the magnetic field lines
embedded in the $z$-$x$ plane, where 
$(z,x) = (r \cos \theta , r \sin \theta )$,
both in the Minkowskian case and in the general relativistic case.
As easily seen from this figure, the magnetic field lines are
moderately modified by the general relativistic effect. 
Owing to this change, curvature radii of the field lines
are also modified by the general relativistic effect.

Mathematically, the radius is defined as
\begin{equation}
 \tilde{\rho} = \left( \frac{d\theta}{dl} \right)^{-1}, 
\end{equation}
where $l$ denotes the proper distance along a field line.
In the Minkowskian case, the field line is simply
specified as
$ \tilde{c}  r = \sin^2 \theta $,
so that the curvature radius 
along the line labeled by $\tilde{c}  $
is given by
\begin{equation}
 \tilde{\rho} = \frac{\sin \theta}{\tilde{c}} 
   \sqrt{1+3 \cos^2 \theta} .
\end{equation}
The general relativistic counterpart 
should be obtained numerically.
\begin{figure}
  \epsfxsize = 8cm
  \centerline{\epsfbox{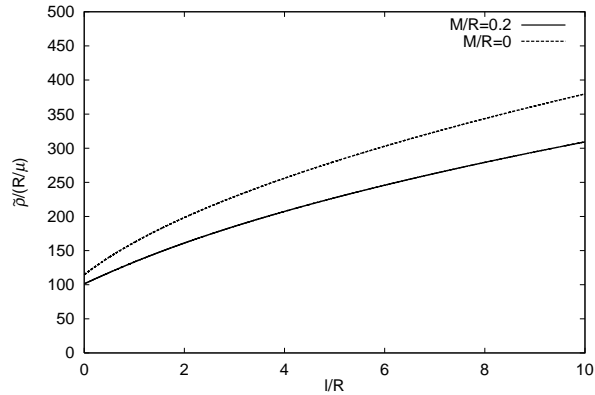}}
  \caption{Curvature radii of magnetic field lines that flow
       from the stellar surface with $\theta = 1^{\circ}$.
       The radii are plotted as functions of the proper distance $l$
       along the field line. The solid line denotes the general
       relativistic case with $M/R=0.2$, while the dashed line 
       denotes the Minkowskian case $M/R=0$. The radii 
       $\tilde{\rho}$ and the proper distance $l$ are 
       normalized by $R/\mu$ and $R$, respectively.}
  \label{fig2}
\end{figure}
Figure \ref{fig2} displays the curvature radii 
$\tilde{\rho}$ of magnetic field lines which
start from the stellar surface with an angle of 
$\theta=1^{\circ}$. 
From Fig.~\ref{fig2}, we find that the general relativistic
effect causes the curvature radius to become
smaller for a fixed magnetic moment.
The  curvature radiation is  produced by 
charged particles moving along the magnetic field lines.
The resulting curvature radiation 
photon energy is proportional to $\tilde{\rho}^{-1}.$
The correct treatment in curved space-time implies
an increase of the photon energy.
Although we have displayed only one comparison between 
the flat and curved cases, 
almost the same results were obtained 
for all small values of $\theta$.
\begin{figure}
  \epsfxsize = 8cm
  \centerline{\epsfbox{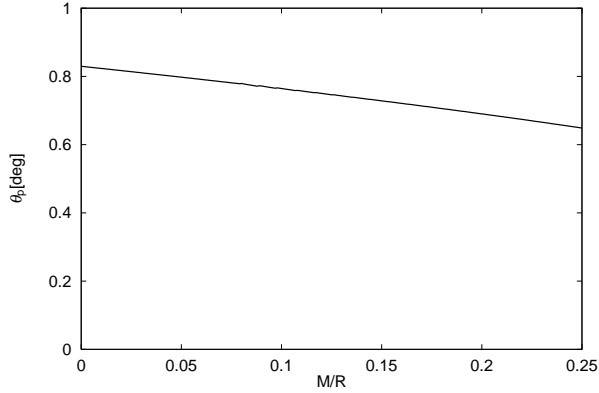}}
  \caption{Polar cap angles plotted as functions of the 
      general relativistic factor $M/R$ 
      for $R_{L} \simeq 5\times10^{3} R$.}
  \label{fig3}
\end{figure}

A modification of the field lines, further, leads to a
change of the polar cap radius. 
The polar cap angle  $\theta_{p}$ is given by 
\begin{equation}
 \theta_{p} = \sin^{-1} 
   \sqrt{\frac{a_{\phi}(R_{L})}{a_{\phi}(R)}},
\end{equation}
where $R_{L}$ is the radius of the light cylinder.
To derive the polar cap angle explicitly, we have assumed
\begin{equation}
 R_{L} = \frac{c}{\Omega} \simeq 5 \times 10^{3} R
\end{equation}
for any value of $M/R$. 
Figure \ref{fig3} displays the
dependence of the polar cap angle $\theta_{p}$ on the
general relativistic factor $M/R$.
From this figure, we see that the polar cap angle is 
reduced by about 15\% due to the curved nature.

\section{Discussion}
Recent observations of compact stars have
given remarkable results that demand the refinement of
theoretical models.
Inspired by this, we have reconsidered an exterior electromagnetic
field surrounding a rotating star endowed with an aligned dipole
magnetic field in the context of general relativity.
The electromagnetic fields were derived in analytic and 
approximate forms.
We found that the expressions calculated in
the flat space-time are accurate within a factor of 
approximately 2.
We have not calculated the 
emission and propagation of radiation
from polar caps of pulsars, but rather have
discussed the implications for the underlining physical processes.
We have found that the general relativistic effects 
increase the strength of electric fields and
decrease the curvature radii of the magnetic field lines. 
Both of these factors contribute to increase the photon 
energy emitted from charged particles.  
The magnitude of the correction 
is of order $ M/R. $
Another important general relativistic effect,
which has not been considered here, is the redshift factor.
The observed energy is shifted to a lower value by 
a factor of $ M/R. $
It is not clear whether or not all 
these general relativistic effects are canceled.
It is important to
construct detailed models of pulsar radiation, 
taking these factors into account.

 Although we have restricted our investigation to a rotating 
star in a vacuum, it seems that actual neutron stars
are surrounded by plasma. Hence, it is important to
investigate the acceleration of charged particles and the 
radiation taking into account the plasma distributions
around stars. A general relativistic analysis
using a certain pulsar model which specifies the plasma 
distribution has been given by 
Muslimov and Tsygan.\cite{mt2}
The general relativistic effects in
pulsar models are not yet clear, since
the magnitude of the effects significantly depends
on the plasma distribution.
It is necessary to discuss the effects
in a more general framework.
This will be the subject of future investigation.

\section*{Acknowledgements}
This work was supported in part by a Grant-in-Aid for 
Scientific Research Fellowship of the Ministry of 
Education, Science, Sports and Culture of Japan 
(No.~12001146).



\end{document}